\documentclass[twocolumn,aps,showpacs,nofootinbib,prd]{revtex4-1}
\usepackage{amssymb}
\usepackage{amsmath}
\usepackage{float}
\usepackage[dvips]{color}
\usepackage{graphicx}
\usepackage{tensor}
\usepackage{braket}
\usepackage{xcolor}
\usepackage{float}
\usepackage{color}
\usepackage{soul}
\usepackage[colorlinks,citecolor=blue,linkcolor=blue]{hyperref}

\newcommand{\be}{\begin{equation}}
\newcommand{\ee}{\end{equation}}
\newcommand{\bea}{\begin{eqnarray}}
\newcommand{\eea}{\end{eqnarray}}
\newcommand{\beas}{\begin{eqnarray*}}
\newcommand{\eeas}{\end{eqnarray*}}

\newcommand{\nn}{\nonumber}

\begin{document}
\title{Magnetic catalysis of a finite size pion condensate}
\author{Alejandro Ayala$^{1,2}$, Pedro Mercado$^1$, C. Villavicencio$^3$}
\affiliation{$^1$Instituto de Ciencias
  Nucleares, Universidad Nacional Aut\'onoma de M\'exico, Apartado
  Postal 70-543, M\'exico Distrito Federal 04510,
  Mexico.\\
  $^2$Centre for Theoretical and Mathematical Physics, and Department of Physics,
  University of Cape Town, Rondebosch 7700, South Africa.\\
  $^3$Departamento de Ciencias B\'asicas, Facultad de Cienicas, Universidad del B\'io-B\'io,
  Casilla 447, Chill\'an, Chile.}

\begin{abstract}

We study the Bose-Einstein condensation of a finite size pion gas subject to the influence of a magnetic field.  We find the expressions for the critical chemical potential and temperature for the onset of condensation. We show that for values of the external magnetic flux larger than the elemental flux, the critical temperature is larger than the one obtained by considering only finite size effects. We use experimentally reported values of pion source sizes and multiplicities at LHC energies to show that if the magnetic flux, produced initially in peripheral heavy-ion collision, is at least partially preserved up to the hadronic phase, the combined finite size and magnetic field effects give rise to a critical temperature above the kinetic freeze-out temperature. We discuss the implications for the evolution of the pion system created in relativistic heavy-ion collisions.
 
\end{abstract}


\maketitle

\section{Introduction}\label{I}

Bose-Einstein condensation (BEC) of a pion system has been vastly explored, mainly as a possible state of matter present in compact stars~\cite{chargedBEC}. The possibility to produce BEC in a relativistic pion system at high temperature has also been a long-studied and sought-after phenomenon. The copious production of these particles in relativistic heavy-ion reactions is one of the primary motivations both to study the problem~\cite{highT-BEC, Begun:2015ifa, Begun:2008hq} and to look for signals~\cite{BE-correlations1}. Among these, one of the most promising seems to be the search for non fully chaotic behavior in interferometry studies~\cite{BE-correlations2}. 

Since condensation is a low momentum phenomenon, in the past, some studies have searched for signatures in the low $p_t$ region of the pion spectrum~\cite{pT}. However, the spectrum at low $p_t$ also contains the contribution of long-lived hadronic resonances, which tends to obscure any possible signal. 

Another interesting avenue to study BEC considers the finite size effects of the pion system produced in heavy-ion reactions. It has been shown that the high $p_t$ part of the spectrum widens for a finite size system~\cite{AS}. The effect is due to the Heisenberg relations, whereby to a finite uncertainty in the transverse location of the particle corresponds a larger spread of its momentum. The treatment used the discrete eigenstates for relativistic scalars subject to hard sphere boundary conditions. However, since the spectrum at high $p_t$ gets also enhanced by flow effects, possible signals due to finite sizes tend to be obscured as well. 

An alternative look into the study of finite size effects is to search for the consequences on the critical temperature $T_c$ for the onset of condensation. If $T_c$ turns out to be in between the kinetic freeze-out $T_{th}$ and the chemical freeze-out $T_{chem}$ temperatures, the possibility that the pion source produces a partially coherent component, increases. It has been shown~\cite{AS, Begun:2008hq} that for systems of the size thought to be formed in heavy-ion collisions, finite volume effects produce a moderate increase of the critical temperature for pion condensation. When in this approach the volume is taken to infinity (the thermodynamic limit), one recovers the usual expressions describing the onset of condensation and the critical temperature for a relativistic pion gas. The approach lends itself to the study of event-by-event fluctuations of conserved charges where it becomes crucial to carefully distinguish the system's from the heat bath's volume. Indeed, in order to observe grand canonical fluctuations of the conserved charges, the ratio of the total conserved charge carried by the system to that by the bath should be a small number~\cite{CCNS}.

In recent years it has also been realized that in peripheral heavy-ion collisions, a large magnetic field is produced affecting the interaction region and thus the pion system created in the aftermath of the reaction. Several calculations show that, at the very early stages of the collision, the field strength can be as high as several times the pion mass squared. However, the field strength is a fast decreasing function of time. Depending on the centrality and collision energy, by the time hadronization happens, the field strength decreases to a fraction of the pion mass squared~\cite{mag}.

It is then natural to ask what, if any, is the effect of these fields on the condensation process. As is well known, in a weakly interacting system, the presence of a magnetic field catalyses condensation. This means that the process is aided by the magnetic field which translates into a larger critical temperature for condensation. The question we set up to answer is whether this can also be the case when dealing with a relativistic pion gas with a charged component in the presence of a magnetic field~\cite{B-BEC,ourB-BEC}. As we show in this work, the answer is positive. We find that when the magnetic flux, passing through the volume formed by the pion system, is larger than the elemental flux, pion condensation is catalyzed.

The work is organized as follows: In Sec.~\ref{II} we set up the formalism to obtain the expressions for the critical chemical potential in the presence of an external magnetic flux in a finite-size pion gas. We work in the weak field limit and show that in this case, the magnetic field affects mainly the ground state. In Sec.~\ref{III} we compute the critical temperature as a function of the particle density. We show that the magnetic field contributes to increase the critical temperature and that the effect goes on top of that produced by a finite volume. We use experimental pion multiplicities and fireball sizes to test whether the enhanced critical temperature results in temperatures close to kinetic freeze-out conditions in heavy-ion collisions. We summarize and discuss the implications of the results in Sec.~\ref{concl} and leave for the appendix the calculation of the excited states' contribution to the particle density in the presence of the external magnetic field, showing that in the weak field limit, such contribution is ${\mathcal{O}}(eB)^2$, where $eB$ is the intensity of the magnetic field, and are also suppressed by extra inverse powers of the volume, and thus that it can be ignored when the volume is finite but considered as large.

\section{Condensation conditions in a finite volume}\label{II}

Consider a non-interacting gas of pions at finite temperature $T=1/\beta$. In the absence of magnetic field effects, the occupation number of a state labeled by the pion's momentum $\vec{p}$ is given by
\bea
\braket{n_{p,j}}=\frac{1}{\mathrm{exp}\left[\beta\left(\sqrt{p^2+m^2}-\mu_j \right)\right]-1},
\label{occupation}
\eea
where $m$ is the pion mass, $p=|\vec{p}|$ is the magnitude of the pion momentum and $\mu_j$, with $j=+,-,0$, represents the chemical potential corresponding to the positive, negative and neutral pions, respectively. The chemical potentials are given by
\bea
   \mu_+&=&\mu + \mu_Q\nn\\
   \mu_-&=&\mu - \mu_Q\nn\\
   \mu_0&=&\mu,
\label{chempot}
\eea
where $\mu$ is the chemical potential associated to the average pion number and $\mu_Q$ is the chemical potential associated to the electric charge. Hereafter we take the pion system as overall neutral, which means that the total number of positive and negative pions is the same, therefore $\mu_Q=0$. 

Let us now consider the same system of pions, this time subject to the influence of a uniform magnetic field directed along the $\hat{z}$ direction, $\vec{B}=B\hat{z}$. The energy levels occupied by the charged pions become discrete Landau levels and the occupation number becomes
\bea
\!\!\!\!\braket{n_{p_z}^\pm}=\frac{1}{\mathrm{exp}\left[\beta\left(\sqrt{p_{z}^2+m^2+(2l+1)eB}-\mu\right)\right]-1},
\label{charged}
\eea
where $e$ is the absolute value of the pion's charge, $p_z$ is the pion momentum along the $\hat{z}$ axis and $l\geq 0$ labels the Landau level.  The neutral pions are still distributed according to Eq.~(\ref{occupation}). 

The particle density can be written as
\bea
   \!\!\!\!\rho(T,\mu;B)&=& \rho_0(T,\mu;B) +  \rho_+(T,\mu;B) +  \rho_-(T,\mu;B)\nn\\
   &=& \frac{1}{V}\sum_{\vec{p}}
   \frac{1}{e^{\beta\left(\sqrt{p^2+m^2}-\mu \right)}-1}\nn\\
   &+& 2\frac{eB}{2\pi L} 
   \sum_{p_z,l}\frac{1}{e^{\beta\left(\sqrt{p_z^2+m^2+(2l + 1)eB}-\mu \right)}-1},
\label{density}
\eea
where we have used that the expression for the number of states for a given Landau level is the same for positive and negative pions. Also, we have written that the length scale $L$ and the volume $V$ are related by $V=L^3$. The factor $eB/2\pi$ accounts for the (uniform) density of states for a given Landau level. 

In order to study the condensation conditions, we need to separate the ground, or condensate state, $\rho^0(T,\mu;B)$ contribution from the excited states $\rho^*(T,\mu;B)$ contribution to the particle density, namely
\bea
    \rho(T,\mu;B)=\rho^0(T,\mu;B) + \rho^*(T,\mu;B),
\label{groundplusexcited}
\eea
where 
\bea
    \rho^0(T,\mu;B) &=& \frac{1}{V}
    \frac{1}{e^{\beta\left(m-\mu \right)}-1}\nn\\
    &+& 2\frac{eB}{2\pi L}\frac{1}{e^{\beta\left(\sqrt{m^2+eB}-\mu \right)}-1},
\label{ground}
\eea
and
\bea
     \!\!\rho^*(T,\mu;B) &=& \int_0^{\infty} \!\!\!\!
    \frac{d^3p}{(2\pi)^3}\frac{1}{e^{\beta\left(\sqrt{p^2+m^2}-\mu \right)}-1}\nn\\
    &+& \frac{2eB}{(2\pi)^2}\sum_{l=1}^{\infty}
    \int_0^{\infty} \!\!\!\!\frac{dp_z}{e^{\beta\left(\sqrt{p_{z}^2+m^2+(2l+1)eB}-\mu \right)}-1}.\nn\\
\label{excited}
\eea
In writing Eq.~(\ref{excited}) we have taken the continuum limit for the sum over momenta and extended the lower limit of integration to zero, since, as shown in Ref.~\cite{Begun:2008hq}, the particle number density of the first of the momentum excited states is smaller than the ground state contribution for large, though finite, $V$. The first line on the right-hand side of Eq.~(\ref{ground}) corresponds to the ground state contribution of the neutral pions, whereas the second line comes from the ground state contribution of the charged pions.

Let us first concentrate on the ground state contributions. Introducing the expression for the elemental flux
\bea
   \Phi_0\equiv \frac{hc}{e}=\frac{2\pi}{e},
\label{elementalflux}
\eea
and expressing the magnetic flux going through the pion system as
\bea
   \Phi=V^{2/3}B,
\label{flux}
\eea
one gets
\bea
   \rho^0(T,\mu;B) &=& \frac{1}{V}\frac{1}{e^{\beta\left(m-\mu \right)}-1}\nn\\
   &+&
   2\frac{\Phi}{\Phi_0}\frac{1}{V} \frac{1}{e^{\beta\left(\sqrt{m^2+eB}-\mu \right)}-1}.
\label{intermsofflux}
\end{eqnarray}
Let us now work in the weak field limit, that is, in the case where $eB< mT$. We thus can write 
\bea
   \frac{\sqrt{m^2+eB}}{T}\approx \frac{m}{T}+\frac{eB}{2mT},
\label{aprox1}
\eea
and therefore
\bea
    \rho^0(T,\mu;B) =
    \frac{1}{V}\frac{1}{e^{\delta}-1}
    +  2\frac{\Phi}{\Phi_0}\frac{1}{V} \frac{1}{e^{\delta+\frac{eB}{2mT}}-1},
    \label{aprox2}
\eea
where we defined $\delta \equiv \beta\left(m-\mu\right)$. Let us now work in the situation where  $\delta > eB/2mT$.  On the other hand, when the system is close to condensation, we also have $\delta \ll 1$. Expanding to leading order in $\delta$ and in the weak field limit, Eq.~(\ref{aprox2}) becomes
\bea
    \rho^0(T,\mu;B) \simeq \frac{1}{V\delta} + 2\frac{\Phi}{\Phi_0}\frac{1}{V \delta},
\label{aprox3}
\eea
where we have discarded terms of order $eB(\Phi/\Phi_0)/V \sim (\Phi/\Phi_0)^2/V^{5/3}$, which are suppressed by extra inverse powers of the system's volume. In writing Eq.~(\ref{aprox3}) we are setting the problem to work in the so called \lq\lq$\Phi$-scheme\rq\rq, that is, where the flux, and not the field intensity alone, is what matters throughout the evolution of the system.

Let us now look at the contribution from the excited states, Eq.~(\ref{excited}). As we show in the appendix, when working in the weak field limit, we can use the Euler-Maclaurin expansion to perform the sum over the Landau levels, with the result
\bea
    &&2\frac{2eB}{(2\pi)^2}\sum_{l=1}^{\infty}
    \int_0^{\infty} \!\!\!\!\frac{dp_z}{e^{\beta\left(\sqrt{p_{z}^2+m^2+(2l+1)eB}-\mu \right)}-1} =\nn\\ 
    &&2 \int\dfrac{d^3p}
    {(2\pi)^3}
   \frac{1}{e^{\beta\left(\sqrt{p^2+m^2}-\mu \right)}-1} - \frac{(eB)^2}{48\pi^2}\sqrt{\frac{2}{mT\delta^3}}.
\label{afterEM}
\eea
When expressing the second term in Eq.~(\ref{afterEM}) as proportional to $(\Phi/\Phi_0)^2$, its contribution is also suppressed by extra inverse powers of the system's volume and becomes sub leading. Therefore, the contribution from the excited states coming from the charged pions, in the weak field limit, is equivalent to the contribution from two neutral pions. Overall, Eq.~(\ref{excited}) can be approximately written as
\bea
   \rho^*(T,\mu) \simeq 3 \int\dfrac{d^3p}{(2\pi)^3}
   \frac{1}{e^{\beta\left(\sqrt{p^2+m^2}-\mu \right)}-1}.
\label{aprox4}
\eea
Notice that in this approximation, the effect of the magnetic field for the description of the condensation conditions comes only from the ground state. 

We now follow Ref.~\cite{Begun:2008hq} to find an expression for Eq.~(\ref{aprox4}) in the limit where $\delta\ll 1$. An approximate expression for Eq.~(\ref{aprox4}) is obtained by looking at the difference between $\rho^*(T,\mu)$ and $\rho^*(T,m)$, that is, the density in excited states and the same object evaluated at $\mu = m$
\bea
   \!\!\!\!\!\!&&\rho^*(T,m) - \rho^*(T,\mu) = \frac{3}{2\pi^2}\int_0^\infty dp p^2\nn\\
   &&\times
   \left\{
   \frac{1}{e^{\beta\left( \sqrt{p^2 + m^2} - m \right)}-1} - \frac{1}{e^{\beta\left( \sqrt{p^2 + m^2} - \mu \right)}-1}
   \right\}.
\label{diffexc}
\eea 
To carry out the integration in Eq.~(\ref{diffexc}) we use the procedure described in Ref.~\cite{Begun:2015ifa}, changing the variable 
\bea
   p=x\sqrt{(2m)(m-\mu)},
\label{changevar}
\eea
and the identity
\bea
   \frac{1}{\exp (y) - 1}=\frac{\coth (y/2) - 1}{2}.
\label{identity}
\eea
Expanding for small $\delta$ and using the series expansion
\bea
   \coth (ay) \sim \frac{1}{ay} + \frac{ay}{3},
\label{seriesex}
\eea
valid for small $y$, and the value of the integral
\bea
   \int_0^\infty dx x^2 \left( \frac{1}{x^2} - \frac{1}{x^2 + 1}\right) = \frac{\pi}{2},
\label{integral}
\eea
we obtain
\bea
    \rho^*(T,\mu) \simeq  \rho^*(T,m) - \frac{3(mT)^{3/2}}{\sqrt{2}\pi}\delta^{1/2}. 
\label{denstityexc}
\eea
Therefore, combining Eqs.~(\ref{aprox3}) and~(\ref{denstityexc}), the expression for the density of states to leading order in the magnetic field and the system's volume, close to the condensation transition, can be written as
\bea
   \rho(T,\mu)&=&\rho^*(T,m)-\frac{3}{\sqrt{2}\pi}(mT)^{3/2}\delta^{1/2}\nn\\
   &+&\frac{1}{V}\left( 1 + 2\frac{\Phi}{\Phi_0}\right)\delta^{-1}.
   \label{eqfordelta}
\eea
Upon defining
\bea
   a(T)&=&\frac{(mT)^{3/2}}{\sqrt{2}\pi},\nn\\
   b(T)&=&\frac{\rho(T,\mu) - \rho^*(T,m)}{3},\nn\\
   c&=&\frac{1}{3}\left( 1 + 2\frac{\Phi}{\Phi_0}\right),
\label{abc}
\eea
and multiplying Eq.~(\ref{eqfordelta}) by $\delta V/3$, we obtain an algebraic equation for $\delta$ written as
\bea
   a(T) V \delta^{3/2} + b(T)V\delta - c=0.
\label{deltaeq}
\eea
We seek the solution to Eq~(\ref{deltaeq}) for $\delta$ in the limit where $V$ is large but finite for the cases where $T$ is $(i)$ less, $(ii)$ equal or $(iii)$ larger than $T_c$.

$(i)$ $T<T_c$: In this case, when the temperature becomes small, we can neglect $a(T)$ and the solution is
\bea
   \delta_{T<T_c} = \frac{c}{b(T)V}.
\label{casei}
\eea

$(ii)$ $T=T_c$: In this case, right at the onset of condensation, we can consider that the contribution to the particle density comes  only from the excited states in which case we can tale $b(T)=0$ and the solution becomes
\bea
   \delta_{T=T_c} = \left( \frac{1+2\frac{\Phi}{\Phi_0}}{3a(T)V}\right)^{2/3}.
\label{caseii}
\eea

$(iii)$ $T>T_c$: In this case the total density comes from excited states and the quantity $\rho^*(T,m)$ is larger than $\rho(T,\mu)$, since the former is evaluated at the maximum value of $\mu$. This means that $b(T)<0$ and the solution becomes
\bea
   \delta_{T>T_c} = \frac{b^2(T)}{a^2(T)}-\frac{1+2\frac{\Phi}{\Phi_0}}{3b(T)V} \simeq \frac{b^2(T)}{a^2(T)}.
\label{caseiii} 
\eea
Notice that since $\Phi$ is bound from below by $\Phi_0$, the limit for the minimum possible magnetic flux is obtained from $\Phi/\Phi_0\to 1^+$. In this limit, the expression for all $\delta$'s, and in particular for $\delta_{T=T_c}$, reduce to the ones obtained in Ref.~\cite{Begun:2015ifa}. For a finite $\Phi$ larger than the elemental flux $\Phi_0$, Eq.~(\ref{caseii}) shows that the critical value of the chemical potential $\mu_c^\Phi$ is smaller than the corresponding value $\mu_c$ in the absence of the magnetic field, and that the former is given by
\bea
   \mu_c^\Phi=m-\frac{1}{m}\left(\frac{2\pi^2}{V^2}\right)^{1/3}
   \left(\frac{1+2\frac{\Phi}{\Phi_0}}{3}\right)^{2/3}.
\label{mucritcond}
\eea

Let us now explore the consequences for the critical temperature.
\begin{figure}[t]
\begin{center}
\includegraphics[scale=0.65]{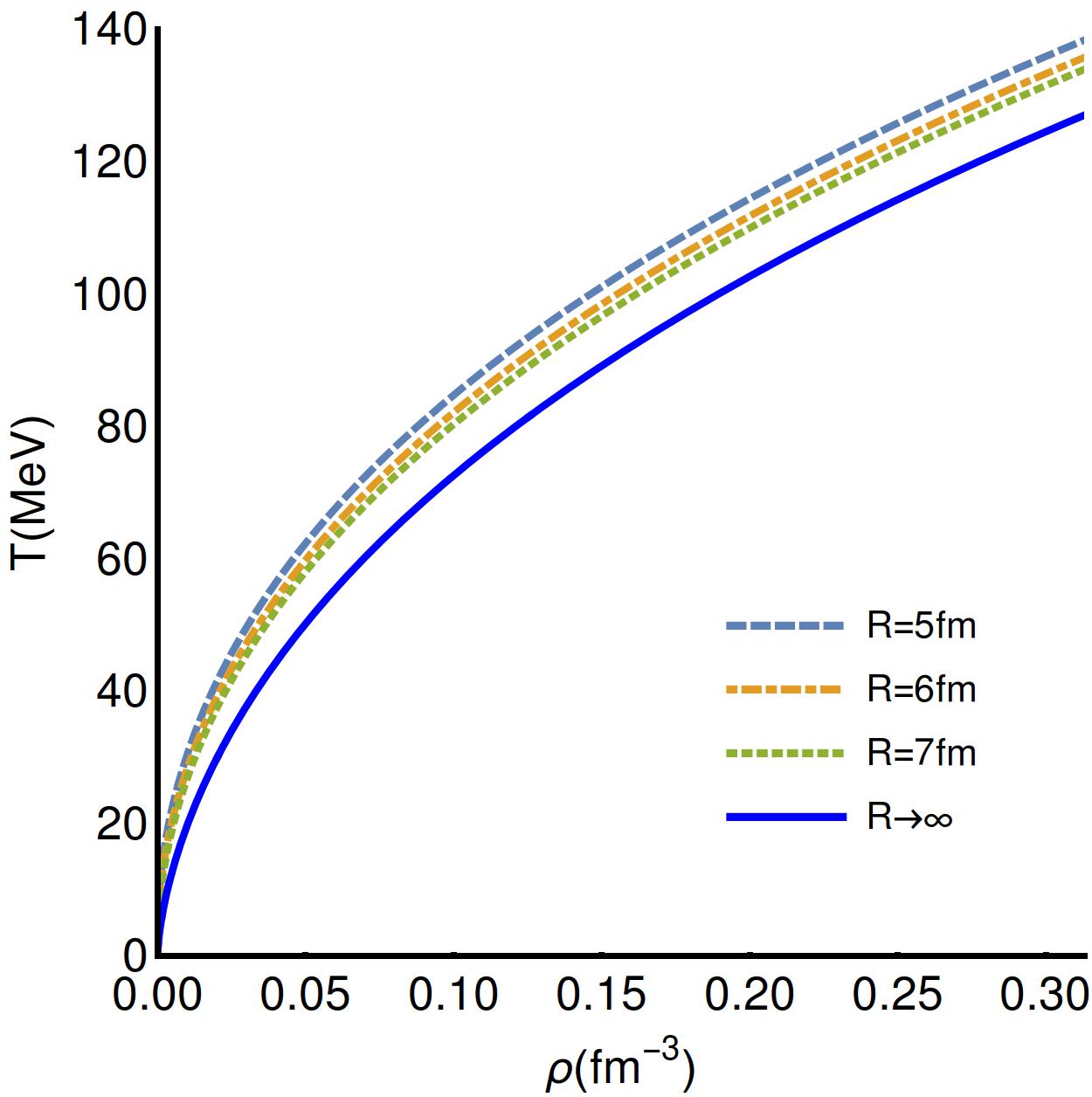}
\end{center}
\caption{Critical temperature for BEC as a function of the system's density in the absence of magnetic field effects. For a given value of the density, the critical temperature increases as the system's size decreases. For comparison we also show the case where the volume is taken to infinity.}
\label{fig1}
\end{figure}

\section{Critical temperature}\label{III}

\begin{figure}[t]
\begin{center}
\includegraphics[scale=0.65]{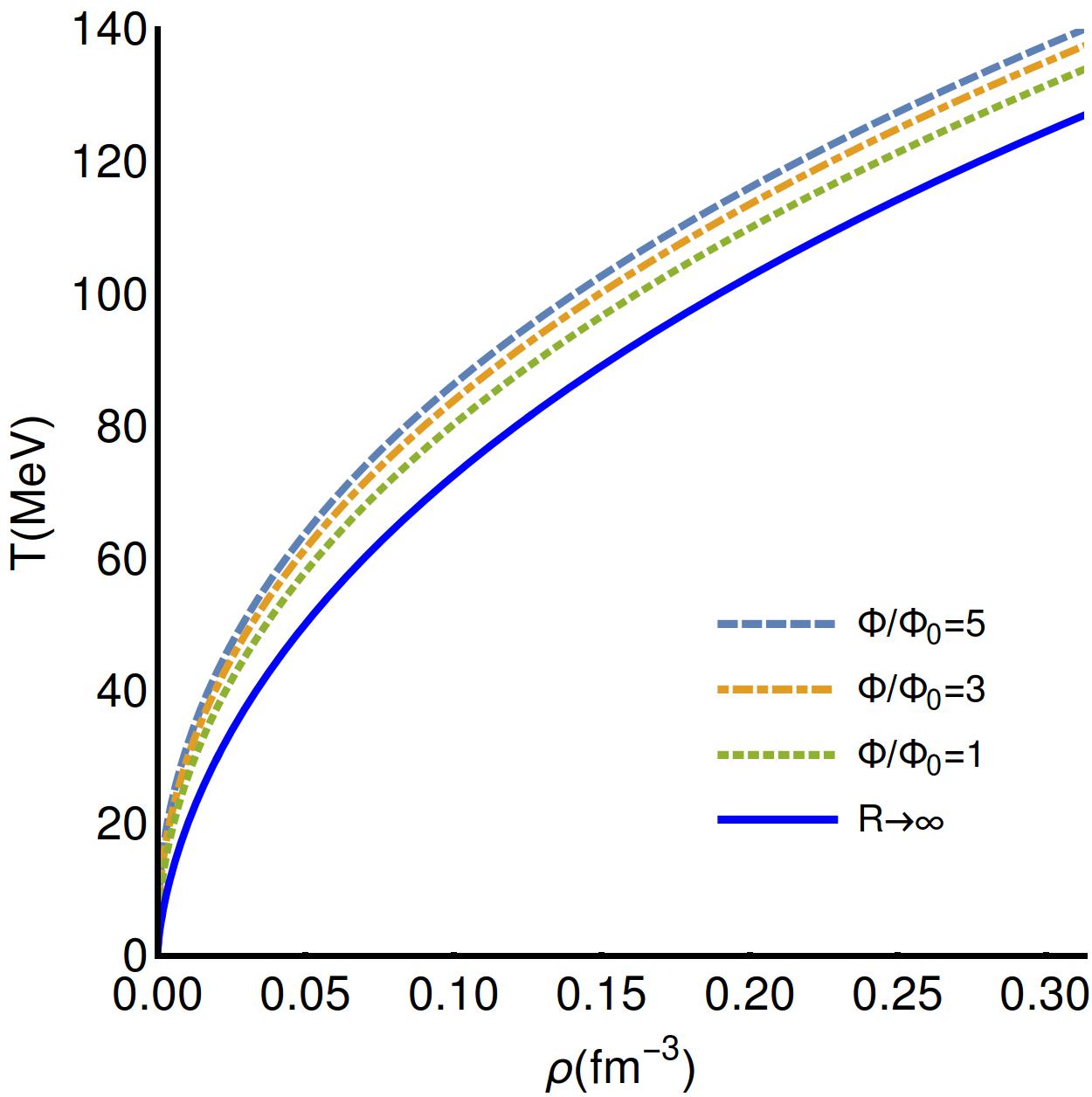}
\end{center}
\caption{Critical temperature for BEC as a function of the density for a fixed system's radius $R=7$ fm and several values of the magnetic flux. For a given value of the density, the critical temperature increases as the magnetic flux increases. For comparison we also show the case where the volume is taken to infinity in the absence of a magnetic field.}
\label{fig2}
\end{figure}

When $T=T_c$, the density comes from the contribution of the excited states, that is $\rho(T_c,\mu_c;B)=\rho^*(T_c,\mu_c;B)$. The density for the excited states can be approximated in the continuum limit, at leading order in the magnetic field, as
\bea
    \!\!\!\!\!\rho^*(T,\mu)&=&3\int_0^{\infty} \!\!\!\!
    \frac{d^3p}{(2\pi)^3}\frac{1}{e^{\beta\left(\sqrt{p^2+m^2}-\mu \right)}-1}\nn\\
    &=&\frac{3Tm^2}{2\pi^2}\sum_{n=1}^{\infty} \frac{1}{n} 
    K_2\left(nm/T\right)\mathrm{exp}\left(n\mu/T\right),
\label{densitycrit}
\eea
where the second line on the right-hand side comes from expanding the denominator of the integrand in a geometrical series, such that the integration over the momentum can be carried out term by term~\cite{AS}. At criticality, we have
\bea
   \rho(T_c,\mu_c^\Phi)=\frac{3T_cm^2}{2\pi^2}\sum_{n=1}^{\infty} \frac{1}{n} 
    K_2\left(nm/T_c\right)\mathrm{exp}\left(n\mu_c^\Phi/T_c\right),\nn\\
\label{critcond}
\eea
where for $\mu_c^\Phi$ we use Eq.~(\ref{mucritcond}). Notice that in the thermodynamic limit, the condition for the critical chemical potential becomes just $\mu_c =m$, irrespective of the magnetic flux. Equation~(\ref{critcond}) gives the critical temperature $T_c$ as a function of the particle density for different ratios $\Phi/\Phi_0$. The solution is shown in Figs.~\ref{fig1} and~\ref{fig2}. Figure~\ref{fig1} shows the effect of varying the size of the system for the case where no magnetic field effects are present, namely the case $\Phi/\Phi_0=1$. Notice that for a given value of $\rho$, the critical temperature increases as the system's volume decreases. For comparison we also show the case where the volume is taken to infinity.  Figure~\ref{fig2} shows the effects of varying the ratio $\Phi/\Phi_0$ for a fixed volume. Notice that for a given value of $\rho$, the critical temperature increases as the flux of the external magnetic field increases~\cite{ourB-BEC}. This means that the magnetic field catalyzes the onset of condensation. For comparison we again show the case where the volume is taken to infinity in the absence of magnetic field effects.

We now proceed to use current experimental data on pion multiplicities and source sizes as well as commonly accepted values for magnetic field strengths to test whether these numbers produce a critical temperature close to reported values for the kinetic freeze-out temperature.

Let us first estimate values of the parameter $\Phi/\Phi_0$ inferred from heavy-ion collisions. For simplicity, let us assume that the system's volume is a sphere of radius $R$. In peripheral heavy-ion collisions, where magnetic fields are produced, the initial overlap region is better described as having an \lq\lq almond" shape. Nevertheless, for our purposes, where what matters is the transverse (to the direction of the magnetic field) area, the description in terms of a sphere suffices. From Eqs.~(\ref{elementalflux}) and~(\ref{flux}), and expressing the field intensity as a multiple $\zeta$ of the pion mass squared, namely, $eB=\zeta m^2$, we can write
\bea
   \frac{\Phi}{\Phi_0}&=&\left(\frac{\pi R^2}{2\pi}\right)(eB)\nn\\
   &\simeq& \zeta\left(\frac{1}{4}\right)(R\ ({\mbox{fm}})\ )^2,
\label{ratioexpl}
\eea
where we have expressed the factor $m^2$ in units of inverse fm$^2$ using that 1 fm $\simeq (1/197)$ MeV$^{-1}$, and $R$ is to be given in fm. In a peripheral collision of symmetric systems, the semi-minor axis of the almond-shape interaction region ($R$ for our purposes) is related to the nuclear radius $R_N$ and to the impact parameter $b$ by
\bea
   R=R_N-b/2.
\label{relation}
\eea
For example, for semi-central collisions, taking $b=R_N$, we see that $R=R_N/2$. For heavy ions, $R_N\sim$ 6 - 7 fm. Therefore, a magnetic field whose initial strength is given by $\zeta\sim 1/2$ is able to produce fluxes with $\Phi/\Phi_0>1$. In high energy heavy-ion collisions, magnetic fields of intensity $\zeta\sim 1$ - 10 in early stages, have been estimated~\cite{mag}. Therefore $\Phi/\Phi_0$ can definitely be larger than one at the beginning of the reaction. If the magnetic flux does not decrease much throughout the system's evolution, up to the hadronic phase, these intense initial fields can leave an imprint on the final hadronic spectra. Whether an initial flux is at least partially preserved or not depends on the transport properties of the plasma produced after the reaction, a subject of intense research over the last years. For the purposes of this work, we assume that the flux can approximately be preserved, though we use conservative values for the ratio $\Phi/\Phi_0$.
Next, we estimate the density of the pion system. We use the simple model of Ref.~\cite{BF} where one can read out that to a semi-central collision with $b=R_N\sim 7$ fm corresponds a centrality of about 30\%. From the data on Ref.~\cite{radii}, which corresponds to Pb + Pb collisions at $\sqrt{s_{NN}}=2.76$ TeV, we can read that the acceptance corrected average number of charged pions within a pseudo rapidity interval $|\eta|< 0.8$ is $\langle N^{ch}\rangle = 426$. It has been estimated that about 1/2 of these pions come from decays of long-lived hadronic resonances~\cite{hadronization}.  Pions from these resonances can hardly be considered thermal and therefore they cannot be included for the determination of the critical temperature. We thus correct the above number taking out the contribution of these resonances and estimate the number of thermal charged pions as $\langle N^{ch}_{th}\rangle\sim 213$. Considering also that about one third of the total number of pions are neutral, we estimate that for this system and at these energies, the average total number of thermal pions in the femptoscopic source is $\langle N_{th}\rangle\sim 320$. From the same Ref.~\cite{BF} one can also read that to $\langle N^{ch}\rangle = 426$ corresponds a source radius $R\sim 7$ fm. Therefore, the average density of thermal pions comes to be about $\langle\rho\rangle=\langle N_{th}\rangle /(4\pi R^3/3)\sim 0.22$ fm$^{-3}$. Accounting for the lowest and highest number of (acceptance and resonance corrected) pions corresponding to the same centrality, it can also be inferred from Ref.~\cite{BF} that the density of the thermal pion system varies in between $0.2<\rho<0.25$ fm$^{-3}$.
\begin{figure}[t!]
\begin{center}
\includegraphics[scale=0.65]{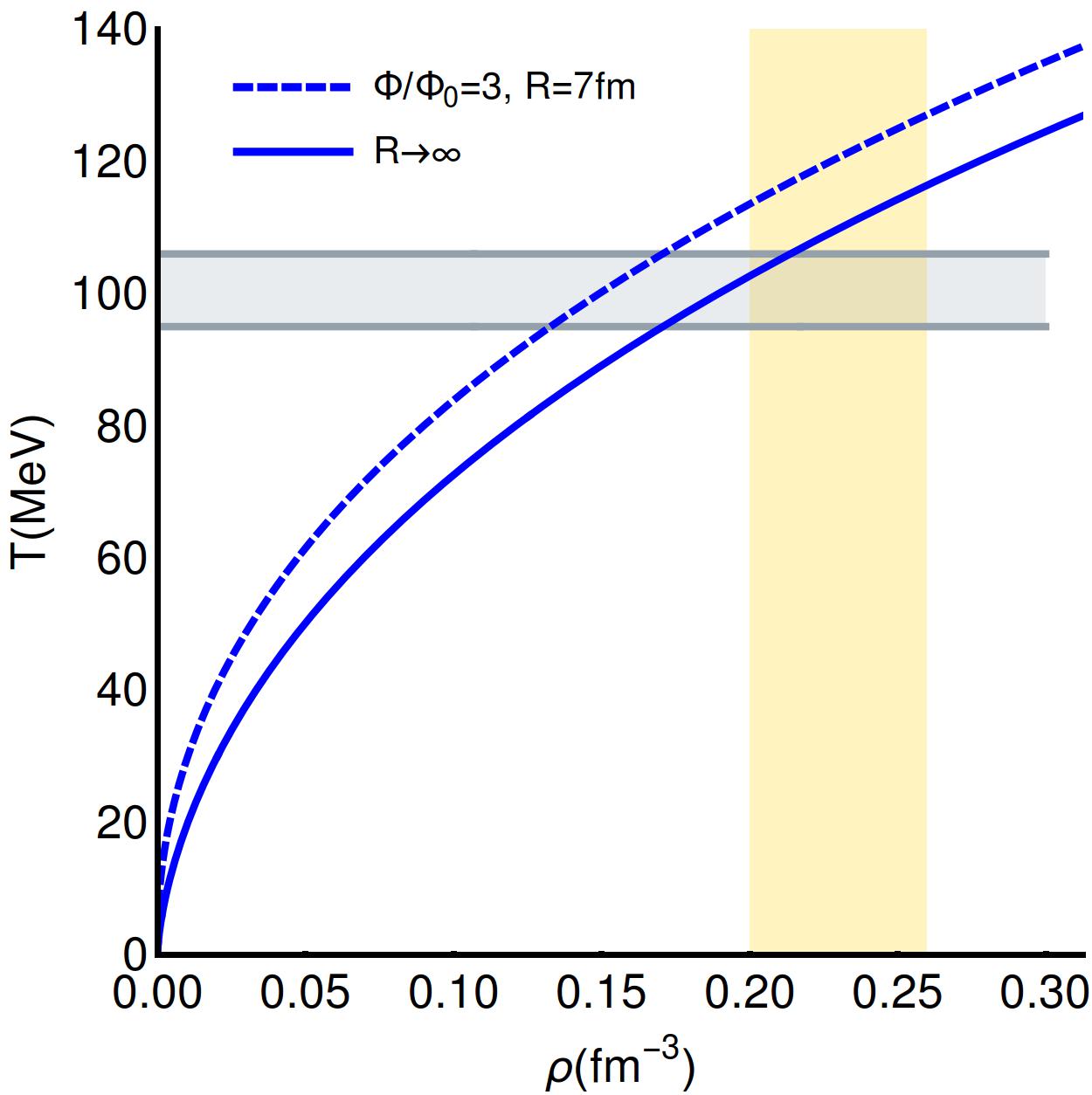}
\end{center}
\caption{Critical temperature for BEC  as a function of the system's density. For comparison we show a range of freeze-out temperatures from central to semi-central collisions and a range of densities and magnetic fluxes in semi-central collisions ($R\sim 7$ fm) at LHC energies. Notice that even for moderate magnetic fluxes the critical temperatures obtained are above the freeze-out temperature range.}
\label{fig3}
\end{figure}
Figure~\ref{fig3} shows the range of critical temperatures corresponding to the range of densities achieved by the pion system created in semi-central heavy-ion collisions at LHC energies for different values of the ratio $\Phi/\Phi_0$. Shown is also the range of pion thermal freeze-out temperatures $95<T_{th}<106$ MeV, reported in Ref.~\cite{ALICE-Tth} and obtained from a blast-wave fit to the pion low $p_t$ distributions. The range corresponds to centrality classes between 0-40\%. Note that even in the absence of a magnetic field ($\Phi/\Phi_0=1$), the thermal freeze-out temperatures are slightly below the values for the condensation critical temperature. Also, even a moderate magnetic flux makes $T_c$ to further increase with respect to $T_{th}$. 

\section{Summary and discussion}\label{concl}

In this work we have found the BEC conditions for a relativistic pion gas occupying a finite volume and subject to the influence of a magnetic field. The field effects are encoded in the flux $\Phi$ across the pion system, referred to the elemental flux $\Phi_0$. We have shown that values of $\Phi/\Phi_0>1$ can be achieved in heavy-ion collisions when the intensity of the field at the beginning of the reaction is higher than half the pion mass squared. Also, under the assumption that this flux is still such that $\Phi/\Phi_0>1$ up to the hadronic phase, we have shown that the critical temperature for pion BEC increases above the increase caused by finite size effects. Using experimentally reported values for the system's size and multiplicities, we have shown that this critical temperature is above the thermal freeze-out temperature extracted from the pion spectra. We emphasize that the chemical potential we work with corresponds to the particle density and not to the baryon density. Although at top energies at RHIC and at the LHC the baryon chemical potential is low, the particle number density is high, and this translates into a high pion density. Nevertheless, since the field strength is a fast decreasing function of time, one expects that by the time pions are formed, the strength of the magnetic field has already decreased such that the highest of the energy scales is the temperature with the magnetic field strength becoming the smallest of these energy scales, which justifies the weak field approximation we use.

The picture that emerges for the evolution of the thermal component of the pion system created in high energy heavy-ion collisions is as follows: At the initial stages of a peripheral collision, a strong magnetic field is generated. This field permeates through the system formed in the reaction zone. The evolution of matter proceeds to chemical freeze-out where particle abundances are established and continues evolving toward thermal freeze-out. However, if the matter generated is able to sustain at least part of the initially crated magnetic flux throughout the evolution up to the hadronic a phase, the finite size pion system meets conditions appropriate for a part of it to occupy the condensate state before it finally reaches kinetic freeze out.  Such partial transit through the condensate state can leave an imprint for example on the chaoticity parameter measured in interferometry studies~\cite{BE-correlations2,radii}, where a partially coherent emission is not discarded.

Notice that in this work we use the approximation where the pion system is treated as a pion gas, namely, we neglect interactions that change, in particular, the pion number. Thus, although the magnetic field breaks isospin invariance, in this approximation its sole effect is its influence on the pion density in the condensed state. What we have shown is that for a given temperature, the critical (density) chemical potential decreases from the value it would have had in the absence of the magnetic field, thus aiding condensation. Also, notice that in an interacting theory that accounts for charge conservation, one needs to describe the evolution of the isospin chemical potential. It has been shown in Ref.~\cite{Endrodi} that for the isospin chemical potential, the effect is the opposite: when a condensate is to be produced in a pion system with a charge imbalance, in the presence of a magnetic field, in order to sustain the condensate, one needs to have a higher charge imbalance than in the absence of the field. This happens because one requires more pions (of a given species) to have a denser system in order to overcome the electrical repulsion of the excess, same sign, charge. 

The results of this work are encouraging and open the door for further studies to test the influence of the medium conditions during the evolution of the pion system from hadronization to kinetic freeze out. In particular it is interesting to test whether the interactions of the pion system with the hadronic medium contribute to preserve or else destroy the condensate state and also to study the evolution of the initial magnetic flux from the initial stages to the hadronization phase of the collision. Also interesting is the possibility to study whether the magnetic flux in compact astrophysical objects meets the conditions to aid the formation of a pion condensate and the possible consequences if such is the case. Some of these studies are currently being conducted and will be reported elsewhere.  

\section*{Acknowledgments}

A. A. and P. M. are in debt to E. Cuautle for pointing out to some references and for useful discussions. A. A. is also grateful to A. Kiesel for useful discussions. This work has been supported in part  by  UNAM-DGAPA-PAPIIT grant number IN101515 and by Consejo Nacional de Ciencia y Tecnolog\1a grant number 256494. CV acknowledges support from FONDECYT under grant numbers 1150847, 1130056, 1150471, and
the group {\em F\1sica de Altas Energ\1as} at UBB. 

\section*{Appendix}

Here we show how to compute Eq.~(\ref{afterEM}) in the weak field limit. To simplify the expression, we extend the sum over over Landau levels from the lowest value $l=0$. This can be done since the contribution of the ground state has already been computed separately
\begin{equation}
   2\frac{2eB}{(2\pi)^2}\sum_{l=1}^{\infty}
   \int_0^{\infty} \!\!\!\!\frac{dp_z}{e^{\beta\left(\sqrt{p_{z}^2+m^2+(2l+1)eB}-\mu \right)}-1}.
\label{A1}
\end{equation}
To approximate Eq.~(\ref{A1}), we use the Euler-Maclaurin formula. Consider first the function
\begin{equation}
f(y)=\int_0^{\infty} \!\!\!\!\frac{dp_z}{e^{\beta\left(\sqrt{p_{z}^2+m^2+y}-\mu \right)}-1} .
\end{equation}
Introducing the variable $y$ defined as $y=2leB$, we can write 
\begin{eqnarray}
\label{euler}
\!\!\!\!\!\!\!\int_{y_i}^{y_f}f(y)dy&=&\sum_{l=0}^{j}f(y_i+2leB)\nn\\ 
\!\!\!\!\!\!\!&-&\sum_{k=1}^p \frac{B_k}{k!}\left(f^{k-1}(y_f)-f^{k-1}(y_i)\right)+{\mathcal{R}},
\end{eqnarray}
where $j=(y_f-y_i)/(2eB)$ and ${\mathcal{R}}$ is an error term. The lowest order approximation is obtained from considering $p=2$ and dropping ${\mathcal{R}}$. Recalling that $B_1=1/2$ and $B_2=1/6$, we can rewrite Eq.~(\ref{euler}) as
\begin{eqnarray}
\!\!\!\!\!\!\!\!\!\!\int_{y_i}^{y_f}f(y)dy&=&\left(\frac{1}{2}f(y_i)+f(y_i+2eB)+\ldots +\right. \nonumber \\ 
\!\!\!\!\!\!\!\!\!\! +\ f(y_f-2eB) &+& \left. \frac{1}{2}f(y_f)\right)- \frac{1}{12}\left( f'(y_f)-f'(y_i)\right).
\end{eqnarray}
We introduce the function $g$ defined in terms of $f$ as $f(y)=2eBg(2leB)$. Notice that $f'(y)=(2eB)^2g'(2leB)$ and therefore we can express Eq.~(\ref{euler}) as
\begin{eqnarray}
\label{int}
\int_{y_i}^{y_f}f(y)dy&=&2eB\left(\frac{1}{2}g(y_i)+g(y_i+2eB)+\ldots +\right.\nonumber \\+\ g(y_f-2eB)  &+& \left. \frac{1}{2}g(y_f)\right)- \frac{(2eB)^2}{12}\left( g'(y_f)-g'(y_i)\right) . \nonumber \\
\end{eqnarray}

An analogous calculation using steps of length $eB$ instead of $2eB$ for the sum yields
\begin{eqnarray}
\label{int2}
\int_{y_i}^{y_f}f(y)dy&=&eB\left(\frac{1}{2}g(y_i)+g(y_i+eB)+\ldots +\right.\nonumber \\+g(y_f-eB)  &+& \left. \frac{1}{2}g(y_f)\right)- \frac{(eB)^2}{12}\left( g'(y_f)-g'(y_i)\right) . \nonumber \\
\end{eqnarray}

Now, upon multiplying Eq.~(\ref{int2}) by 2 and subtracting Eq.~(\ref{int}) we notice that all terms of the function $g$ evaluated at even multiples of $eB$ cancel, leaving only terms evaluated at odd multiples of $eB$, namelly
\begin{eqnarray}
\label{int3}
\int_{y_i}^{y_f}f(y)dy&=&2eB\left(g(y_i+eB)+g(y_i+3eB)+\ldots +\right.\nonumber \\ &+&g(y_f-eB)  \left. \right)+ \frac{(2eB)^2}{24}\left( g'(y_f)-g'(y_i)\right).  \nonumber \\
\end{eqnarray}
Setting $y_i=0$ and $y_f\to \infty$ we have
\begin{eqnarray}
&&2\frac{(2eB)}{(2\pi)^2}\sum_{l=0}^\infty g((2l+1)eB)= \nonumber \\ &&2\frac{(2eB)}{(2\pi)^2}\sum_{l=0}^\infty \int_0^\infty \frac{dp_z}{e^{\beta\left(\sqrt{p_{z}^2+m^2+(2l+1)eB}-\mu \right)}-1} =\nonumber \\ &&\frac{1}{2\pi^2}\int_0^\infty f(y)dy -\frac{(eB)^2}{12\pi^2}(g'(\infty)-g'(0)).
\end{eqnarray}
We now identify $y=p_\perp^2$ and form the three-dimensional momentum squared $p^2=p_z^2+p_\perp^2$. We thus can write 
\begin{eqnarray}
\int_0^{\infty} f(y)dy&=&\frac{1}{2}\int_0^{\infty} dy \int_{-\infty}^\infty \frac{dp_z}{e^{\beta\left(\sqrt{p_{z}^2+m^2+y}-\mu \right)}-1} \nonumber \\
&=&\int_0^{\infty} dp_{\perp} p_\perp\int_{-\infty}^\infty \frac{dp_z}{e^{\beta\left(\sqrt{p_{z}^2+m^2+p_{\perp}^2}-\mu \right)}-1} . \nonumber \\
\end{eqnarray}
Using that $d^3p=p_\perp dp_\perp dp_z d\theta$ we get
\begin{eqnarray}
\int_0^{\infty} f(y)dy=4\pi^2 \int \frac{d^3p}{(2\pi)^3} \frac{1}{e^{\beta\left(\sqrt{p_{z}^2+m^2+p_{\perp}^2}-\mu \right)}-1}, \nonumber \\
\end{eqnarray}
from which it follows that
\begin{eqnarray}
\frac{1}{2\pi^2}\int_0^{\infty} f(y)dy=2 \int \frac{d^3p}{(2\pi)^3} \frac{1}{e^{\beta\left(\sqrt{p_{z}^2+m^2+p_{\perp}^2}-\mu \right)}-1}. \nonumber \\
\end{eqnarray}
On the other hand, notice that
\begin{eqnarray}
g'(y)&=&-2\int_0^\infty \frac{dp_z}{2\pi}\frac{e^{\beta\left(\sqrt{p_{z}^2+m^2+y}-\mu \right)}}{\left(e^{\beta\left(\sqrt{p_{z}^2+m^2+y}-\mu \right)}-1\right)^2}\nonumber \\ &\times &\frac{\beta}{\sqrt{p_z^2+m^2+y}}.
\end{eqnarray}
Therefore, $g'(y=\infty)=0$ and
\begin{eqnarray}
g'(y=0)=-2\int_0^\infty \frac{dp_z}{2\pi}\frac{\beta}{\sqrt{p_z^2+m^2}}&& \nonumber \\ \times\left[\frac{1}{\left(e^{\beta\left(\sqrt{p_{z}^2+m^2}-\mu \right)}-1\right)^2}\right. &+& \nonumber \\ +\left. \frac{1}{e^{\beta\left(\sqrt{p_{z}^2+m^2}-\mu \right)}-1} \right].
\label{gprime0}
\end{eqnarray} 
Using Eqs.~(\ref{changevar}) and~(\ref{identity}) we can rewrite Eq.~(\ref{gprime0}) as
\begin{eqnarray}
g'(y=0)&=&-2\frac{\sqrt{2m}(m-\mu)^{1/2}}{(2\pi)T}\nn\\
&\times&
\int_0^\infty \frac{dx}{\sqrt{m^2+2m(m-\mu)x^2}}\nonumber \\ 
&\times&
\left[\frac{\mathrm{coth}\left(\frac{\beta}{2}(m-\mu)(x^2+1)\right)-1}{2} \right.\nonumber \\  &+& \left. \left(\frac{\mathrm{coth}\left(\frac{\beta}{2}(m-\mu)(x^2+1)\right)-1}{2}\right)^2 \right]. \nonumber \\ 
\end{eqnarray} 
Simplifying the expression above, one gets
\begin{eqnarray}
g'(y=0)=-\frac{\sqrt{2m}(m-\mu)^{1/2}}{2(2\pi)T}\int_0^\infty \frac{dx}{m\sqrt{1+\frac{2T}{m}\delta x^2}} && \nonumber \\ \times \mathrm{csch}^2\left(\frac{\beta}{2}(m-\mu)(x^2+1)\right). \nonumber \\ 
\end{eqnarray} 

Recall that we are working in the limit $\delta \ll 1$. Therefore, the leading contribution for $g'(y=0)$ can be written as
\begin{eqnarray}
g'(y=0)&=& -\frac{2\sqrt{2mT\delta}}{(2\pi)\delta^2mT}\int_0^{\infty}\dfrac{dx}{\sqrt{1+\frac{2T}{m}\delta x^2}}\frac{1}{(x^2+1)^2},\nn\\
\end{eqnarray} 
from where, after preforming the integral, we get
\begin{equation}
g'(y=0)=-\frac{1}{4}\sqrt{\frac{2}{mT\delta^3}}
\end{equation}
Combining all of these results, we finally obtain Eq.~(\ref{afterEM}).
%

\end{document}